\documentclass[11pt]{asme2ej}

\usepackage{amsmath}
\usepackage{amssymb}
\usepackage{graphicx,indentfirst,subfigure,epsfig}
\usepackage{enumerate}
\usepackage{subfigure}
\usepackage{subfigmat}
 \usepackage{varioref}
 \usepackage{wrapfig}
\usepackage{hyperref}
\usepackage{cleveref}
\usepackage{booktabs}
\usepackage{cite}
\usepackage{soul}
\usepackage{xcolor}
\usepackage[subtle]{savetrees}
\usepackage{setspace}
\usepackage{etoolbox}
\patchcmd{\subfigmatrix}{\hfill}{\hspace{0.8cm}}{}{}

\usepackage{bm}

  \def\clap#1{\hbox to 0pt{\hss#1\hss}}

\providecommand{\mat}[1]{\bm{#1}}%
\renewcommand{\vec}[1]{\mathbf{#1}}

\providecommand{\mC}{\ensuremath{\mat{C}}}
\providecommand{\mD}{\ensuremath{\mat{D}}}

\providecommand{\mH}{\ensuremath{\mat{H}}}

\providecommand{\mJ}{\ensuremath{\mat{J}}}

\providecommand{\mQ}{\ensuremath{\mat{Q}}}
\providecommand{\mR}{\ensuremath{\mat{R}}}
\providecommand{\mS}{\ensuremath{\mat{S}}}

\providecommand{\mV}{\ensuremath{\mat{V}}}
\providecommand{\mW}{\ensuremath{\mat{W}}}

\providecommand{\mLambda}{\ensuremath{\mat{\Lambda}}}

\providecommand{\vs}{\ensuremath{\vec{s}}}

\providecommand{\vu}{\ensuremath{\vec{u}}}

\providecommand{\vx}{\ensuremath{\vec{x}}}
\providecommand{\vy}{\ensuremath{\vec{y}}}
\providecommand{\vz}{\ensuremath{\vec{z}}}

\makeatletter
\newif\if@checkentries

\makeatother
  
\confyear{2021}

\title{Blade Envelopes Part II: Multiple Objectives and Inverse Design\vspace{-1 cm}}
\author{Chun Yui Wong$^{\dagger}$\thanks{Address all correspondence to Chun Yui Wong, \texttt{cyw28@cam.ac.uk}}, Pranay Seshadri$^{\ddagger \star}$, Ashley Scillitoe$^{\star}$,  Bryn Noel Ubald$^{\star}$\\
\textbf{Andrew B. Duncan}$^{\ddagger \star}$, \textbf{Geoffrey Parks}$^{\dagger}$
    \affiliation{
    $^\dagger$Department of Engineering, University of Cambridge, U.K.\\
    $^\ddagger$Department of Mathematics, Imperial College London, U.K. \\
    $^\star$Data-Centric Engineering, The Alan Turing Institute, U.K.\\
    
    }	
}

\newcommand{\myRe}{\mathbb{R}}

\newcommand{\block}[1]{
  \underbrace{\begin{matrix}\vu_1^T & \vu_2^T & ... & \vu_n^T\end{matrix}}_{#1}
}

\crefname{figure}{Figure}{Figures}
\Crefname{figure}{Figure}{Figures}
\crefname{equation}{}{}
\Crefname{equation}{}{}

\begin{document}

\maketitle    

\begin{abstract}

\textit{Blade envelopes offer a set of data-driven tolerance guidelines for manufactured components based on aerodynamic analysis. In Part I of this two-part paper, a workflow for the formulation of blade envelopes is described and demonstrated. In Part II, this workflow is extended to accommodate multiple objectives. This allows engineers to prescribe manufacturing guidelines that take into account multiple performance criteria. }

\textit{The quality of a manufactured blade can be correlated with features derived from the distribution of primal flow quantities over the surface. We show that these distributions can be accounted for in the blade envelope using vector-valued models derived from discrete surface flow measurements. Our methods result in a set of variables that allows flexible and independent control over multiple flow characteristics and performance metrics, similar in spirit to inverse design methods. The augmentations to the blade envelope workflow presented in this paper are demonstrated on the LS89 turbine blade, focusing on the control of loss, mass flow and the isentropic Mach number distribution. Finally, we demonstrate how blade envelopes can be used to visualize invariant designs by producing a 3D render of the envelope using 3D modelling software.}
\end{abstract}

\section{INTRODUCTION}

In the first part of this two-part paper \cite{wong_blade_nodate}, we defined the concept of a \emph{blade envelope}, leading to a computational guideline yielding automatic scrap-or-use decisions for manufactured turbomachinery components based on models trained from an aerodynamic database. Using the theory of inactive subspaces, a range of geometric designs that are invariant in loss is identified, and geometries from this invariant region can be generated with no additional computational fluid dynamics (CFD) solves. From these geometries, we construct the blade envelope as the distribution of invariant geometries within a control zone. This enables us to calculate the distance of test profiles from the blade envelope using the statistical metric of Mahalanobis distance to gauge whether the profile is expected to give similar-to-nominal performance. Thus, the decision to scrap or keep a measured component reduces down to the computation of this distance from the blade envelope. Throughout this two-part paper, we use the Von Karman Institute LS89 linear turbine cascade \cite{arts_aero-thermal_1992} with conditions as described in Section 2 of Part I for demonstration, but note that the base methodology of blade envelopes extends beyond 2D axial designs.

In this second part, we extend blade envelopes beyond the manufacturing stage of production, and describe how they can be used during the design stage as well.  During the shape design of a highly-loaded turbine stage, the minimization of loss is often accompanied with constraints to avoid trivial solutions where the blade is unloaded. For example, in \cite{torreguitart_optimization_2018}, the exit flow angle is constrained to be above the baseline value to ensure sufficient work extraction. In \cite{montanelli_gradient_2015}, the authors put an equality constraint on the mass flow rate while optimizing the loss coefficient to factor out possible reduction in entropy generation due to reduction in flow capacity. Prior work \cite{seshadri_yuchi_parks_shahpar_2020, seshadri_turbomachinery_2017} has leveraged active subspaces to construct 2D performance maps for compressor blade design. In the latter work, multiple objectives including the pressure ratio and flow capacity are considered by mapping contours of different objectives onto the active subspace of efficiency. Manufacturing deviations are modelled as constant excursions from the nominal design. The main drawback of this approach is the requirement to run further simulations to map out performance contours in the active subspace. In this work, we incorporate multiple aerodynamic design requirements by interpreting them as additional constraints factored into blade envelopes. In upcoming future work, we extend our scope of consideration to structural requirements, and examine its implications on the tolerance covariance as additional constraints. 

In situations where tighter control over the performance of the component is required, constraints on surface flow characteristics can be implemented. Clark \cite{clark_step_2019} establishes the correlation between aerodynamic features---defined via parts of the surface isentropic Mach number distribution---and aerodynamic performance. Control over these key features can be achieved by factoring the isentropic Mach number distribution as an additional \emph{vector-valued} objective in blade envelopes. This approach is similar in spirit to \emph{inverse design}, where a target distribution is specified on the surface of a blade, and the blade shape is iteratively modified to give a geometry that matches the distribution. While inverse design yields an optimal geometry that fits the design criterion over the entire surface, our approach aims to find designs that satisfy the target distribution in locations that are most critical to performance. The relaxation of constraints on other locations allows a range of designs to be specified, whose expanse is explicitly quantified by the blade envelope. Moreover, we can combine the control over the surface flow profile with constraints over other scalar objectives to perform inverse design constrained by requirements on other measures of performance.

\begin{figure}
\centering
\includegraphics[width=0.6\linewidth]{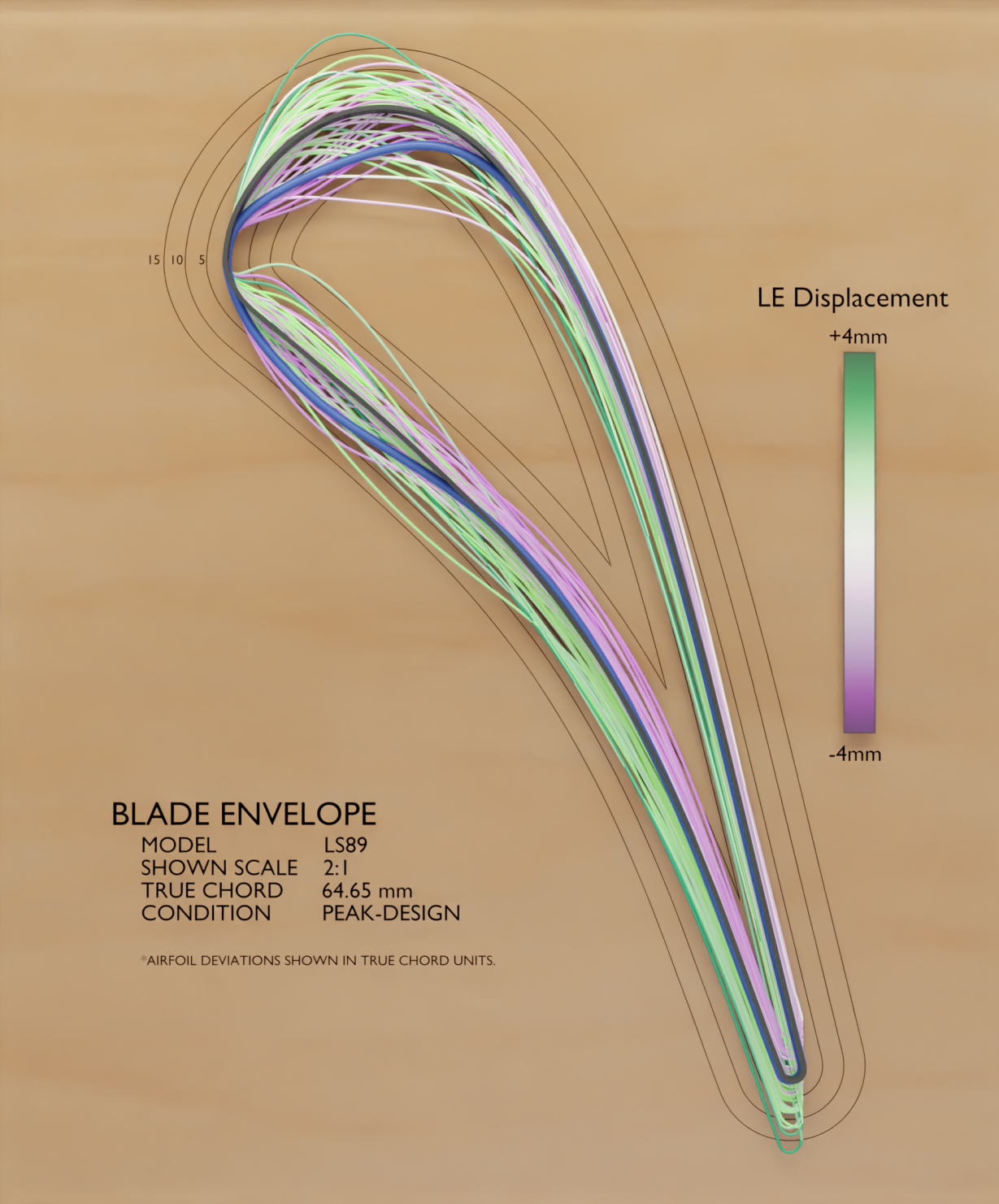}
\caption{A CAD representation of the controlled loss and mass flow rate blade envelope, highlighting one distinct invariant profile in blue. Note: In all 2D and 3D plots in this paper, all deformations are drawn to scale.}
\label{fig:cad_3}
\end{figure}

In addition to computationally quantifying the region of a high-dimensional design space corresponding to designs with invariant objective values, blade envelopes are also amenable to intuitive visualizations. We conclude this work by connecting the numerical output of the blade envelopes workflow to 3D modelling software, producing outputs such as Figure~\ref{fig:cad_3}. The result is a visualization of the possible designs within the blade envelope that can be easily shared between engineers and made interactive via novel technologies such as 3D printing and augmented reality. 

\section{COMPUTATIONAL METHODOLOGIES}

Blade envelopes demarcate boundaries between regions within the space of manufactured geometries which exhibit similar performance metrics, one of which is shown in blue in Figure~\ref{fig:cad_3}.  These envelopes are formed from statistics derived from an aerodynamic database containing geometries sampled from the inactive subspace with respect to a scalar objective. Building upon this framework, we describe two algorithms to identify regions of the design space characterized by invariance in \emph{multiple objectives}, including both scalar- and vector-valued objectives. This enables us to obtain samples invariant in these objectives to form the required database for a multi-objective blade envelope.

Recall from \cite{wong_blade_nodate} the partitioning of the design space into the \emph{active} and \emph{inactive} subspaces via an orthogonal matrix $\mQ = \left[\begin{array}{cc}
\mW & \mV \end{array}\right]$,
\begin{equation}
\vx = \mW \mW^{T} \vx + \mV \mV^{T} \vx,
\end{equation}
where we desire that the quantity of interest varies predominantly with changes in the active coordinates $\mW^T \vx$, and are approximately invariant with respect to changes in the inactive coordinates $\mV^T \vx$. Building on this idea of subspace decomposition, we aim to identify an inactive subspace matrix $\mV$ whose columns form a basis for a subspace within which multiple objectives are approximately invariant.

\subsection{Intersection of inactive subspaces} \label{sec:intersection}

Consider two independent objectives $f_1(\vx)$ and $f_2(\vx)$ depending on the design variables $\vx$ differently. Assuming each objective is differentiable, we can compute the following symmetric positive semi-definite matrices,
\begin{equation}
\label{eqn:grad_cov}
\mC_i = \int_{\mathcal{X}} \nabla_{\vx} f_i(\vx) \nabla_{\vx} f_i(\vx)^T~ \tau~ d\vx, \qquad i=1,~2
\end{equation}
where $\tau = 2^{-d}$ defines a uniform distribution over the design space, which is a $d$-dimensional hypercube $[-1,1]^d$ as before. Their respective eigenvalue decompositions can be evaluated,
\begin{equation} \label{eqn:eig_decomp}
\mC_i = [\mW_i \enskip \mV_i] \begin{bmatrix}
\mLambda_{i,1} & \mathbf{0}\\
\mathbf{0} &\mLambda_{i,2}
\end{bmatrix}
\begin{bmatrix}
\mW_i^T\\
\mV_i^T
\end{bmatrix}, \qquad i=1,~2,
\end{equation}
where the non-negative eigenvalues are ordered in descending
order, and the diagonal terms in $\Lambda_{i,2}$ are sufficiently close to zero, for $i=1,2$.
From \cref{eqn:eig_decomp}, the inactive subspace corresponding to $f_i(\vx)$, i.e., the column span of $\mV_i$, can be obtained. Let's denote this by $\mathcal{V}_i = \text{colspan}(\mV_i)$. The idea is to find the  \emph{intersection} of the inactive subspaces $\mathcal{V}_{int} =  \mathcal{V}_1 \cap \mathcal{V}_2$. If a design perturbation lies within $\mathcal{V}_{int}$, then by definition it lies in both $\mathcal{V}_1$ and $\mathcal{V}_2$, implying that such a perturbation has insignificant effects on both $f_1(\vx)$ and $f_2(\vx)$. How do we find this intersection, and how can we sample from it?

It can be shown that $\mathcal{V}_{int}$ is a linear subspace that can be expressed as the column span of a basis matrix (see \cite[p.~64]{golub2013matrix}). Let this matrix be $\mV_{int}$. Once we can evaluate $\mV_{int}$, the exploration of the subspace $\mathcal{V}_{int}$ subject to constraints reduces to an application of the hit-and-run sampling algorithm as described in Part I \cite{wong_blade_nodate}. To find this matrix, we require the \emph{active} subspace basis matrices $\mW_1 \in \myRe^{d\times r_1}$ and $\mW_2 \in \myRe^{d\times r_2}$, and the assumption that the dimension $r$ of the space $\text{colspan}(\mW_1) \oplus \text{colspan}(\mW_2)$ is less than $d$, the ambient design space dimension, where $\oplus$ denotes the direct sum. With these assumptions, the following recipe can be used to find $\mV_{int}$:
\begin{enumerate}
\item Form $\mW_{both} := [\mW_1 \enskip \mW_2]$.
\item Decompose $\mW_{both}$ by QR-factorization, $\mW_{both} = \mQ \mR$.
\item Set $\mV_{int}$ to be the last $d - r$ columns of $\mQ$.
\end{enumerate}
This gives us the basis matrix $\mV_{int}$ corresponding to the inactive subspace for both $f_1(\vx)$ and $f_2(\vx)$. This procedure is straightforwardly generalizable to $n>2$ objectives via induction, as long as the dimension $r$ of the direct sum of the active subspaces is strictly less than $d$.

\subsection{Vector-valued objectives} \label{sec:vecdr}

Given the method of intersections, a natural extension to find an invariant subspace for a vector-valued objective would be to treat the $N$ components of the objective
\begin{align} \label{eqn:discrete_M}
\begin{split}
\mathbf{M(x)} &= [M_1(\vx),~M_2(\vx),~...,~M_N(\vx)]\\
&= [M_1,~M_2,~...,~M_N]
\end{split}
\end{align}
as independent scalar objectives, and find the intersection of all their respective inactive subspaces $\mathcal{V}_{int} = \mathcal{V}_1 \cap \mathcal{V}_2 \cap ... \cap \mathcal{V}_N$. The challenge with this approach for large $N$ (when compared to $d$) is the fact that the total number of active coordinates is likely to exceed $d$. In this situation, the intersection of the inactive subspaces is trivial, i.e., it contains only the zero vector. An alternative approach that provides approximate invariance across the component objectives is required.

We take cues from the work of Wong et al. \cite{WONG2020113383} and Zahm et al. \cite{zahm_gradient-based_2018} and define the following multivariate generalisation of \eqref{eqn:grad_cov},
\begin{equation}
\mH = \int_{\mathcal{X}} \mJ(\vx)~\mR~\mJ(\vx)^T~ \tau~ d\vx,
\end{equation}
where
\begin{equation}
\mJ \left( \vx \right) = \left[ \frac{\partial M_{1}}{\partial \vx} ,~ \ldots~ , \frac{\partial M_{N}}{\partial \vx} \right] \in \myRe^{d\times N}
\end{equation}
is the Jacobian matrix. The matrix $\mR$ is a user-defined $N\times N$ diagonal \emph{weight matrix},
\begin{equation}
\mR=\begin{bmatrix}
\omega_1 &&&\\
& \omega_2 &&\\
&& \ddots &\\
&&& \omega_N
\end{bmatrix},
\end{equation}
 where $\omega_i \geq 0$ for $1\leq i \leq N$. The corresponding active and inactive subspace basis matrices $\mW_M$ and $\mV_M$ for $\mathbf{M}$ can be found again via the eigenvalue decomposition,
\begin{equation}
\mH = [\mW_M \enskip \mV_M] \begin{bmatrix}
\mLambda_{M,1} & \mathbf{0}\\
\mathbf{0} &\mLambda_{M,2}
\end{bmatrix}
\begin{bmatrix}
\mW_M^T\\
\mV_M^T
\end{bmatrix},
\end{equation}
with $\mW_M \in \myRe^{d\times r}$ and $\mV_M \in \myRe^{d\times (d-r)}$. The dimensionality $r$ is found by identifying the largest eigenvalue gap, as in the scalar case. In \cite{zahm_gradient-based_2018}, the focus is placed on finding the active subspace for vector-valued objectives to achieve dimension reduction. Here, we place the emphasis on the inactive subspace instead, to achieve component-wise invariance in vector-valued quantities.

The power of this method lies in the flexibility in setting the weight matrix $\mR$. It can be shown that $\mH$ is the weighted sum of the gradient covariance matrices corresponding to each component objective \cite[Sec.~5.2.2]{zahm_gradient-based_2018}. That is, if
\begin{equation}
\mC_i = \int_{\mathcal{X}} \nabla_{\vx} M_i(\vx) \nabla_{\vx} M_i(\vx)^T~ \tau~ d\vx,
\end{equation}
then
\begin{equation}
\mH = \sum_{i=1}^N \omega_i \mC_i.
\label{equ:H}
\end{equation}
The larger the weight for $\omega_i$, the bigger its influence on the eigenvectors $\mW_M$ and $\mV_M$.  Samples from the inactive subspace yield output values that are close to nominal at locations with a large prescribed weight; in locations with a smaller weight, output values are allowed to vary more freely. The weights allow us to prescribe different degrees of invariance to various locations indexed by the vector-valued objective.

\subsection{Discussion: Distinguishing between inactive subspaces}

The problem of finding an orthonormal basis for the intersection of multiple inactive subspaces is equivalent to finding the null space of their direct sum (see 6.4.2 in \cite{golub2013matrix}). In other words, there is a connection between the inactive subspace of $\mH$ from \eqref{equ:H}, and last few columns of $\mQ$ in \Cref{sec:intersection}.

Consider the inactive subspace to be the orthogonal complement of the column span of $\mH$, which would correspond to the eigenvectors of $\mH$ that have eigenvalues of zero. In practice, however, if we have many independent objectives, there is a strong possibility that the span of $\mH$ will have full dimension, in which case the inactive subspace will be trivial, i.e., $\left\{ \boldsymbol{0} \right\}$. For practical reasons, this motivates us to consider a relaxed definition of the inactive subspace, which corresponds to the linear span of \emph{small} eigenvalues of $\mH$. The threshold between \emph{small} and \emph{not-small} is problem specific, although one can make various scaling arguments about what a reasonable threshold would be. In this setting, the inactive subspace will depend strongly on the choice of the scaling parameters $\mR$, which can weight the relative importance of difference subspaces.

Utilizing the direct intersection approach in \Cref{sec:intersection} implies that the inactive subspaces of the various objectives are characterized by eigenvalues that are all exactly zero (albeit with a different formulation of the active subspace). We can interpret this as a \emph{strongly} inactive subspace, and it will be used when considering multiple flow quantities. In contrast, the inactive subspace obtained through $\mH$ represents a \emph{weakly} inactive subspace, and is used in this paper for studying a single vector-valued quantity, e.g., isentropic Mach number distribution.

\subsection{Constrained tuning via active coordinates} \label{sec:inverse_design_method}

Assuming that the active subspaces are linearly independent---a mild assumption for independent objectives, using the method of intersection of $n$ scalar- or vector-valued objectives, we obtain a linear decomposition of the design space. This decomposition can be characterised by gathering the active subspace basis matrices together with the intersection of the inactive subspaces,
\begin{equation}
\mD = \begin{bmatrix}
\mW_1 & \mW_2 & ... & \mW_n & \mV
\end{bmatrix},
\end{equation}
yielding a square matrix $\mD \in \myRe^{d\times d}$ with independent columns. Corresponding to $\mD$, a coordinate transformation on the design $\vx$ yields the following
\begin{align}
\begin{split}
\mD^T \vx &= \begin{bmatrix}
\left(\mW_1^T \vx\right)^T & \left(\mW_2^T \vx\right)^T & ... & \left(\mW_n^T \vx\right)^T & \left(\mV^T \vx\right)^T
\end{bmatrix}^T\\
&= \begin{bmatrix}
\smash[b]{\block{r_1+r_2+...+r_n}} & \vz^T\\
\end{bmatrix}^T
\end{split}
\end{align}
The first components of the new coordinates $\vu_1, \vu_2,...,\vu_n$ correspond to the active coordinates of the objectives $f_1, f_2,...,f_n$; the final coordinates $\vz$ correspond to the inactive coordinates. Any perturbation lying in the column space of $\mV$ has its first $r_1+r_2+...+r_n$ coordinates equal to zero, implying that it causes no change to the active coordinates for all objectives. Thus, this perturbation does not cause significant change to any objective.

Assuming that each coordinate is independent, the active coordinates can also be treated as tunable parameters that directly control each objective separately. Adjusting the values of $\vu_1$ while keeping every other active coordinate constant results in changes in $f_1$, while maintaining the values of other objectives. Note that we can allow the inactive coordinates $\vz$ to float, since those coordinates do not affect any objective.

\section{AUGMENTING THE BLADE ENVELOPE FOR LS89}

In the following, we apply the methods described above to extend the blade envelope for the LS89 turbine profile to accommodate a constraint on the mass flow function. Next, the method of conditional tuning is demonstrated by identifying geometries with constant loss and mass flow coefficients by their isentropic Mach number distributions.

\subsection{Inactive subspace for loss and mass flow}

\begin{figure}
\centering
\includegraphics[width=0.6\linewidth]{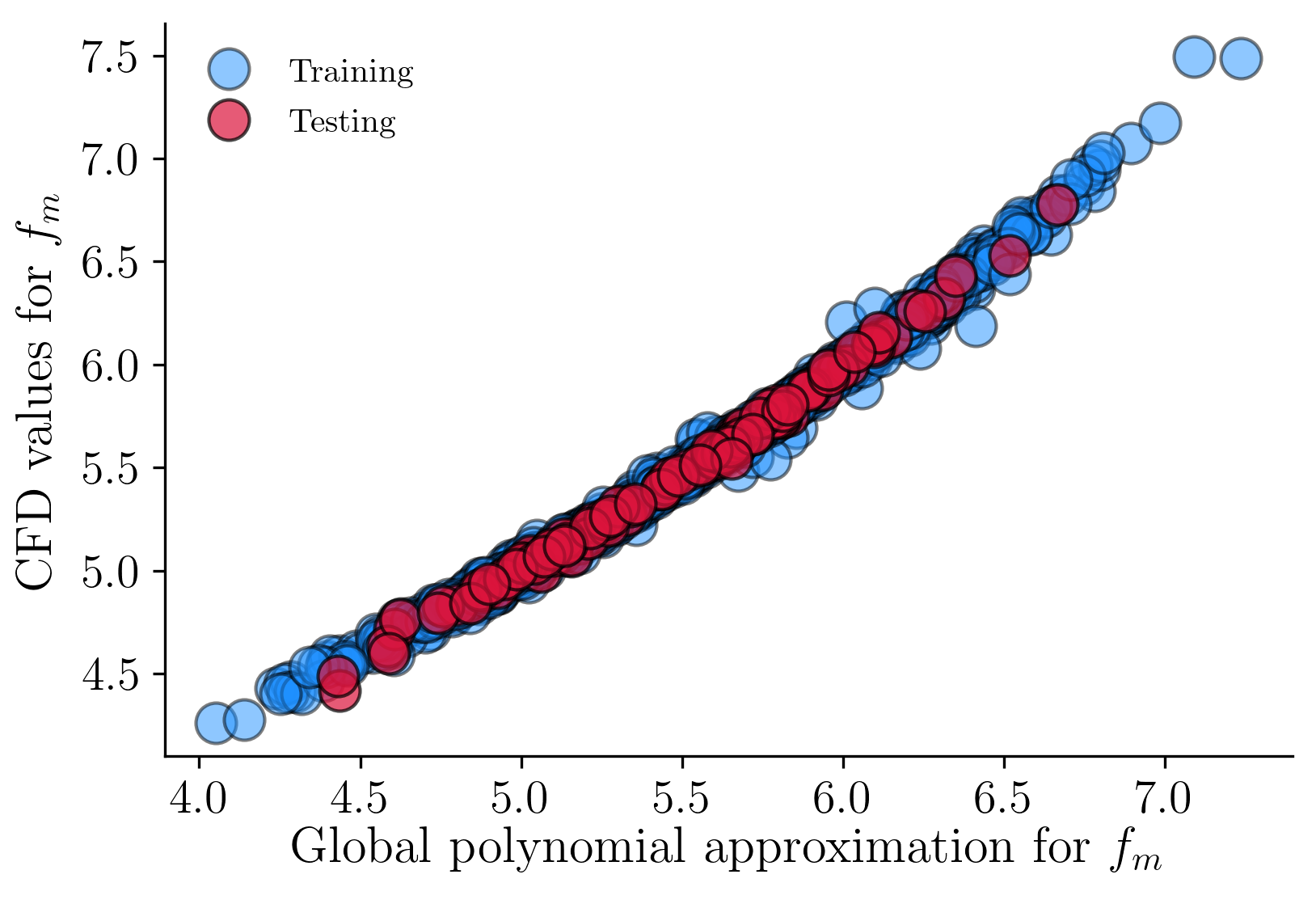}
\caption{Validation of polynomial model for the mass flow function. This coincides with the sufficient summary plot because the global polynomial approximation is a linear model.}
\label{fig:validation_fm}
\end{figure}

Recall from Equation (2) in Part I \cite{wong_blade_nodate} the definition of the mass flow function
\begin{equation}
f_m = \frac{\dot{m} \sqrt{T_{01}}}{p_{01}} \times 10^4.
\end{equation}
Following a similar approach to the calculation of the inactive subspace for loss, we first model the mass flow function using an orthogonal polynomial series via regression on a data-set consisting of $M=800$ training samples and 200 validation samples. These are drawn from a 20D design space based on free-form deformation---the same data-set used in part I. In this case, it is found that a linear polynomial ($p=1$) suffices to yield an $R^2$ value of 0.993 on the validation data (see \cref{fig:validation_fm}). Now, we could compute the associated gradient covariance matrix \eqref{eqn:grad_cov} by sampling gradient evaluations, but since the model is linear, the gradient is constant across the input domain. This  implies that we can read off the 1D active subspace as the linear polynomial coefficients to the design parameters, i.e.,
\begin{equation}
f_m~\approx~u_{f_m} = \mW_{f_m}^T \vx,
\end{equation}
where $\mW_{f_m} \in \myRe^{20\times 1}$ specifies the active subspace, and $u_{f_m}$ is the active coordinate for $f_m$. 

\begin{figure}
\centering
\includegraphics[width=0.6\linewidth]{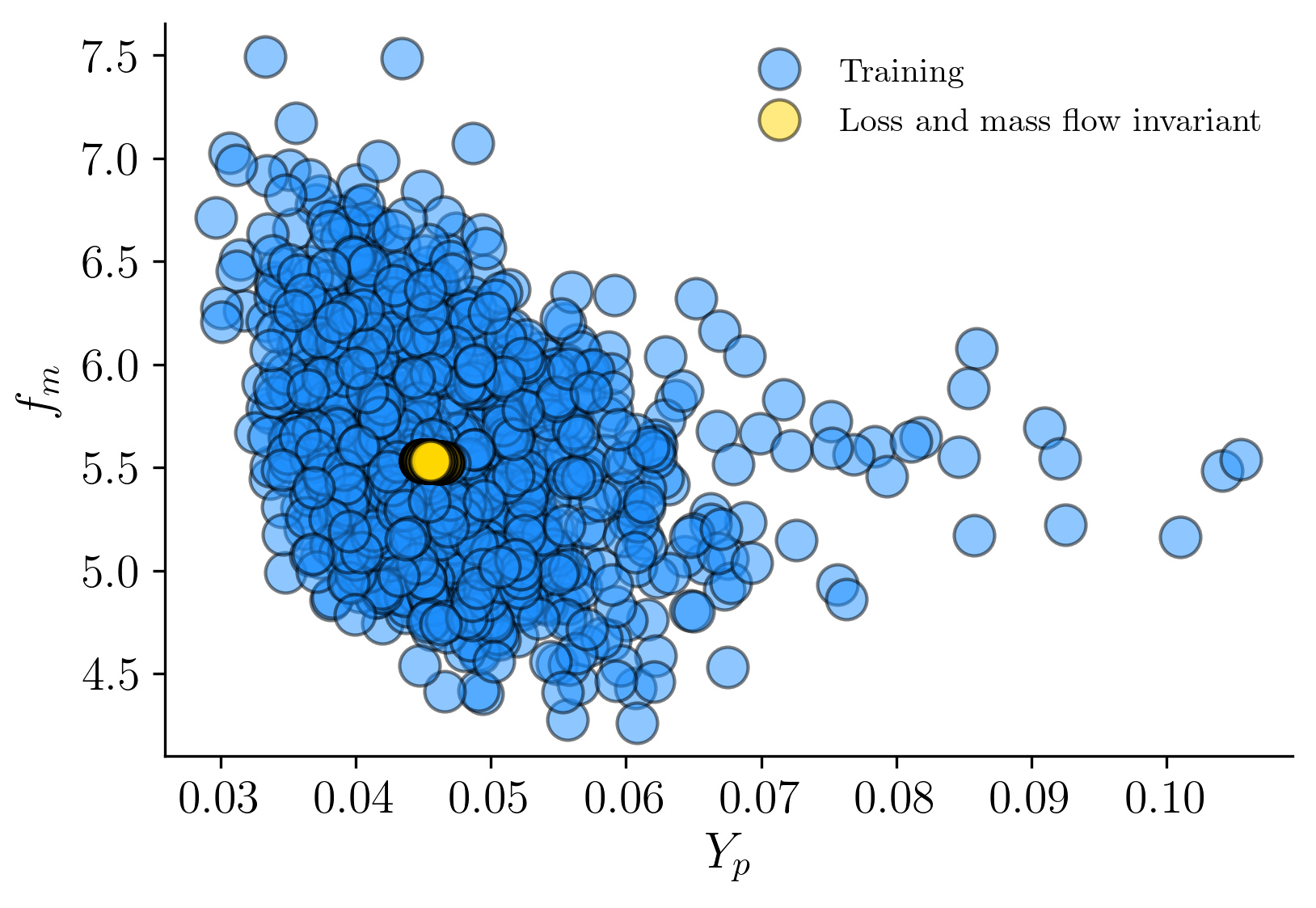}
\caption{Loss and mass flow function for random training designs (blue) and invariant designs within the intersection of the inactive subspaces (yellow).}
\label{fig:invariant_both}
\end{figure}

Equipped with the active subspaces for both loss ($\mW_{Y_p} \in \myRe^{20\times 1}$) and mass flow ($\mW_{f_m}\in \myRe^{20\times 1}$), we can now find the intersection of their inactive subspaces via the method described in \Cref{sec:intersection}. The active subspaces are linearly independent and the intersection $\mathcal{V}_{int}$ is $20-(1+1)=18$-dimensional. Then, using the hit-and-run sampling strategy described in Part I, $H=5000$ samples are generated from $\mathcal{V}_{int}$. The invariance in both objectives is verified by picking 500 designs and running them through the CFD solver to obtain their true loss and mass flow function values, which we plot against the corresponding values for the training data-set in \cref{fig:invariant_both}. It can be seen that the variation in both loss and mass flow is much smaller for invariant designs compared to random ones---the length of the two-standard-deviation interval for $Y_p$ has been reduced from 0.0185 to 0.0011, while that for $f_m$ from 0.847 to 0.008. Note that this figure is not intended to show any correlation between the two objectives, but serves to compare the relative amount of variation in both objectives for training samples and invariant samples respectively. It confirms that the designs in the intersection are approximately invariant in both objectives. 

\begin{figure}
\centering
\includegraphics[width=0.6\linewidth]{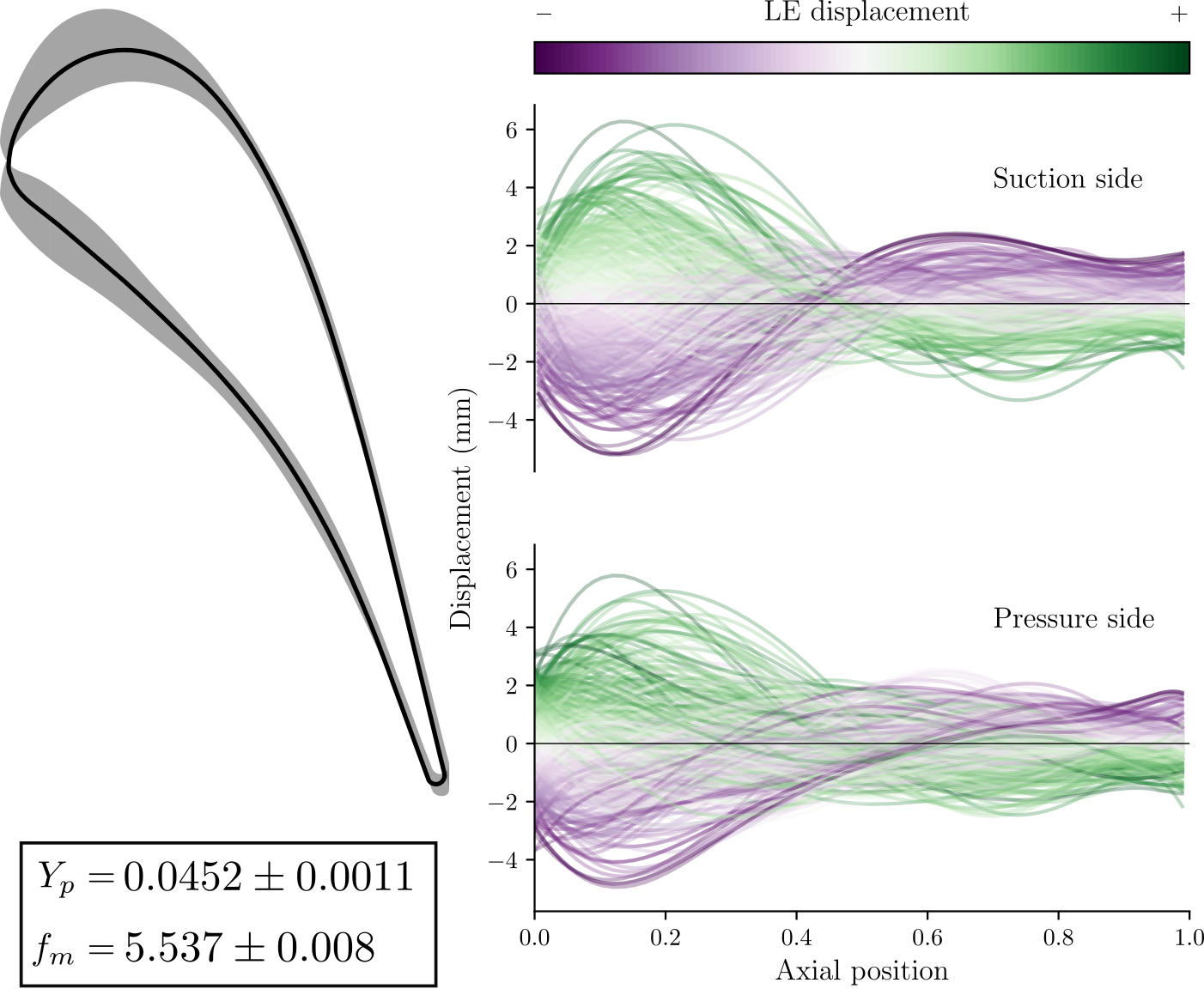}
\caption{The blade envelope for loss and mass flow.}
\label{fig:loss_mass_env}
\end{figure}

\begin{figure}
\centering
\begin{subfigmatrix}{1}
\subfigure[]{\includegraphics[width=0.4\linewidth]{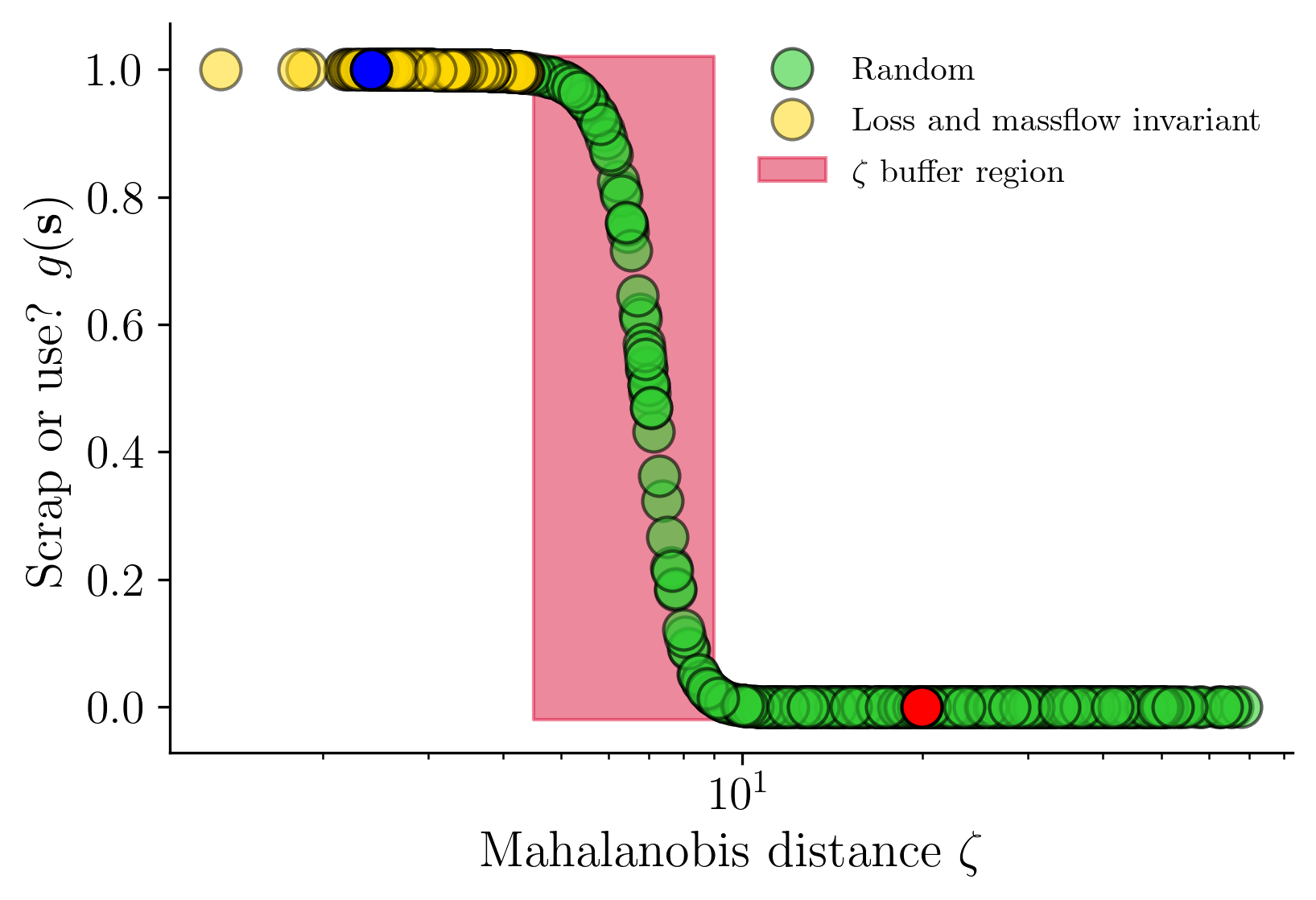}}
\subfigure[]{\includegraphics[width=0.4\linewidth]{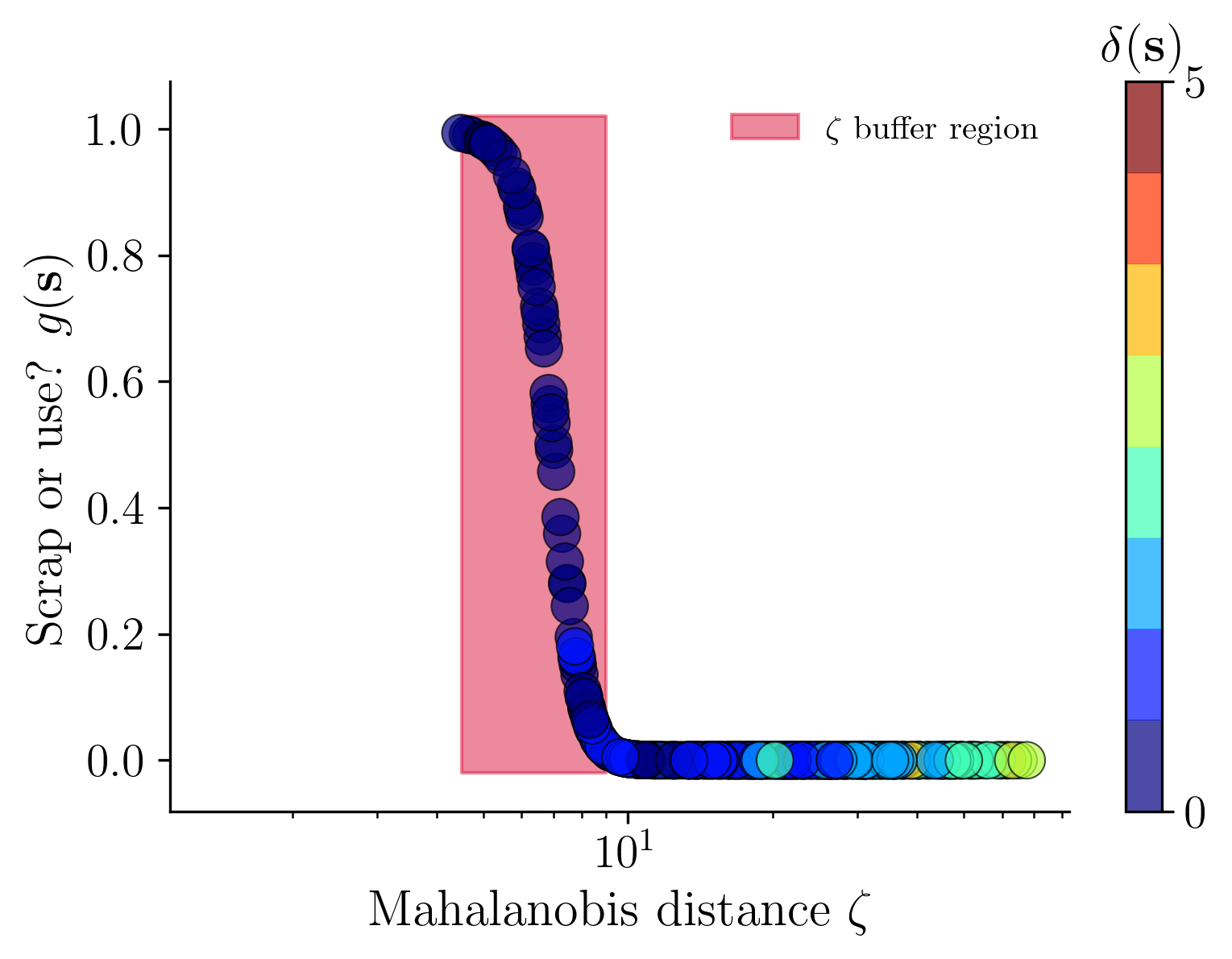}}
\end{subfigmatrix}
\caption{Trained logistic function for applying a binary scrap-or-use decision on profiles, requiring invariance in both the loss and mass flow function; (a) shows samples from the 20D design space and (b) shows samples from the 30D design space described in Section 4.4 of Part I \cite{wong_blade_nodate}. The blue and red geometries are shown in Figure~{\ref{fig:compare}} in detail. In (b), the samples are colored according to their deviation from nominal loss and mass flow, quantified via \cref{eqn:metric}.}
\label{fig:logistic}
\end{figure}

\begin{figure}
\centering
\includegraphics[width=0.5\linewidth]{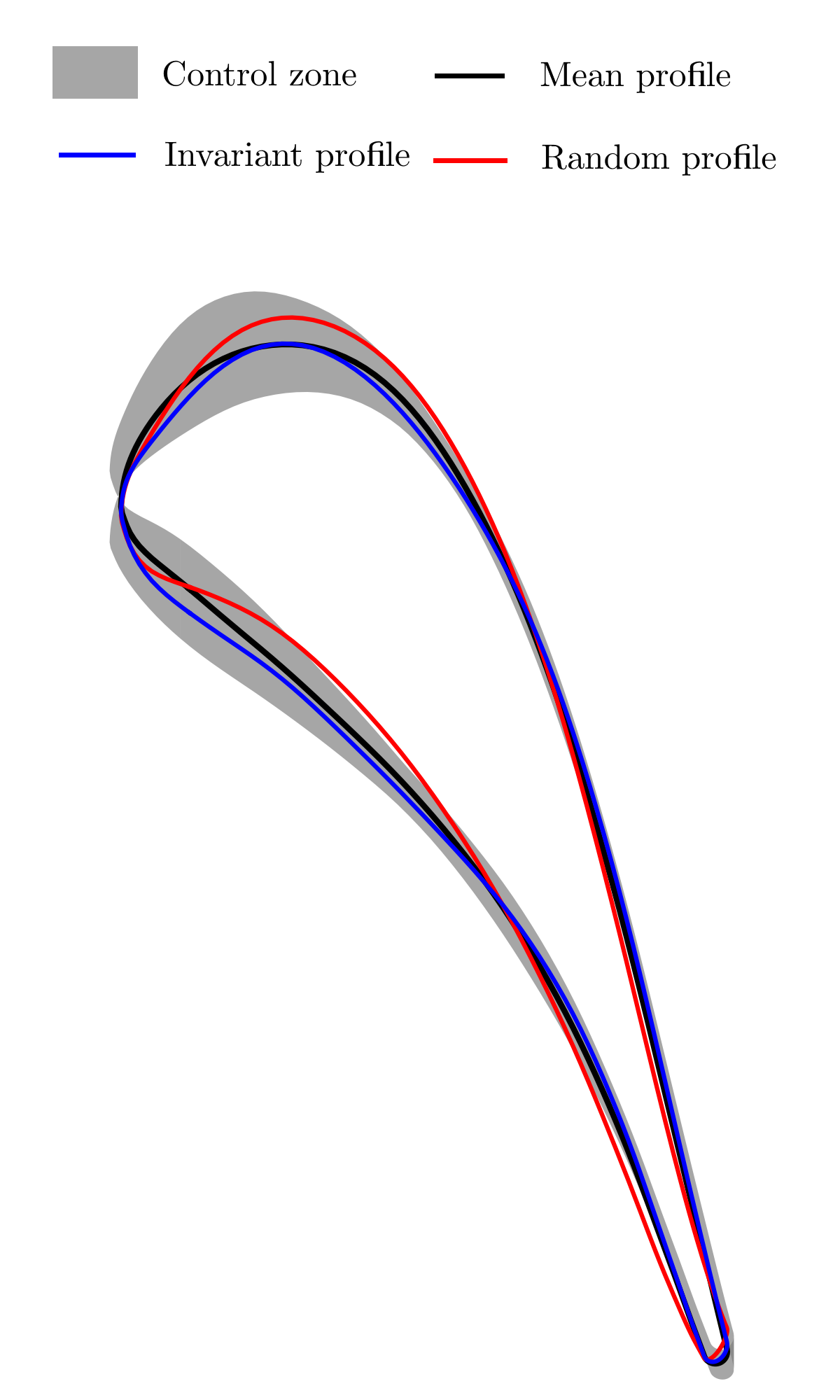}
\caption{Comparing the geometries of two example profiles highlighted in Figure~\ref{fig:logistic}(a).}
\label{fig:compare}
\end{figure}

The blade envelope for invariance in loss and mass flow is shown in \cref{fig:loss_mass_env}, where we have used the LE displacement---the average displacement over the first 5\% of the chord---from Part I to color-code the displacements of invariant profiles to give a visual representation of their typical curvature profiles. Compared to the blade envelope for loss in Part I, the control zone in this case is narrower from mid-chord to the trailing edge because an extra performance constraint on the mass flow function is imposed. The typical curvature of the sample profiles is also milder. Using the sample profiles, the ensemble mean and covariance matrix can be calculated and the Mahalanobis distance can be computed for test profiles, as in Section 3.4 of Part I \cite{wong_blade_nodate}. Recalling equation (23) from Section 3.4 of Part I, the Mahalanobis distance $\zeta(\vs)$ can be substituted into a logistic function with tuned hyperparameters $\beta_1, \beta_2$ and $\beta_3$, 
\begin{equation}
g \left( \vs \right) = \frac{\beta_1}{1 + \text{exp}  \left\{  -\beta_2 \left( \zeta \left( \vs \right) - \beta_3 \right)  \right\}   },
\label{equ:logistic}
\end{equation}
to calculate a scrap-or-use value $g(\vs)$ for a test profile $\vs$. This value indicates our confidence in the test profile satisfying the performance invariance constraints. \Cref{fig:logistic} shows the application of the trained logistic function on profiles from two different design spaces---the 20D design space from which the training data is drawn, and the 30D design space described in Section 4.4 of Part I \cite{wong_blade_nodate}. The samples from the 30D space are colored according to how close their loss and mass flow values are to nominal values calculated according to the following formula
\begin{equation} \label{eqn:metric}
\delta(\vs) = \sqrt{\left(\vy(\vs) - \boldsymbol{\mu}_y \right)^T \mS_y^{-1} \left(\vy(\vs) - \boldsymbol{\mu}_y \right)},
\end{equation}
where $\vy(\vs) = [Y_p(\vs),~f_m(\vs)]^T$ are the objective values for the geometry $\vs$, and $\boldsymbol{\mu}_y$ and $\mS_y$ are the ensemble mean and covariance matrix of $\vy$ over all training designs respectively. Note that $\delta(\vx)$ is another Mahalanobis distance, this time defined over the two output objectives. This is analogous to that defined over geometric profiles (see Section 3.4 of Part I \cite{wong_blade_nodate}). Here, we see similar results as those obtained for a single objective in Sections 4.3 and 4.4 of Part I \cite{wong_blade_nodate}---geometries with low Mahalanobis distance and high scrap-to-use value have low deviation $\delta(\vs)$ from nominal loss and mass flow. This shows that the Mahalanobis distance defined over geometries is an effective diagnostic for identifying profiles invariant in multiple objectives. In Figure~\ref{fig:compare}, two geometries---one close to the blade envelope and one far from it---are shown along with the control zone and mean profile. Although the red profile is mostly inside the control zone, it does not obey invariance in loss and mass flow ($Y_p = 0.072,~f_m = 5.582$) with extra losses incurred due to the sharp variation in displacement on the pressure side. Meanwhile, the blue profile is substantially different from the mean in terms of displacement, but has a similar performance to nominal. Note that in general, neither a comparison of the geometries through visual means or pointwise tolerance approach leads to this result straightforwardly. 

\subsection{Setting the peak isentropic Mach number}
In addition to control over the total quantity of loss, the distribution of this loss is also of interest to the engineer as a means to judge the quality of a component. The invariant subspace of loss and mass flow contains geometries that yield different flow profiles, resulting in a range of loss distributions, some preferred over others. According to Clark \cite{clark_step_2019}, surface flow characteristics defined by primal flow quantities can be used as predictors of aerodynamic performance, differentiating between desirable and undesirable flow profiles. In his work, four features of the surface isentropic Mach number distribution are isolated:
\begin{enumerate}
\item the peak isentropic Mach number,
\item the surface fraction of the peak isentropic Mach number,
\item the leading edge isentropic Mach number, and
\item the pressure side isentropic Mach number prior to acceleration.
\end{enumerate}
In this section, the method of vector-valued dimension reduction is applied to modulate the peak of this distribution. The isentropic Mach number is defined in Equation (3) in Part I \cite{wong_blade_nodate} as
\begin{equation}
M(s) = \sqrt{\frac{2}{\gamma-1} \left(\left(\frac{p_{01}}{p(s)}\right)^{\frac{\gamma - 1}{\gamma}} - 1\right)},
\end{equation}
where $s$ indicates the location on the surface. This can be discretized as in \cref{eqn:discrete_M} into $\mathbf{M(x)}$. 

\begin{figure*}
\centering
\includegraphics[width=0.7\linewidth]{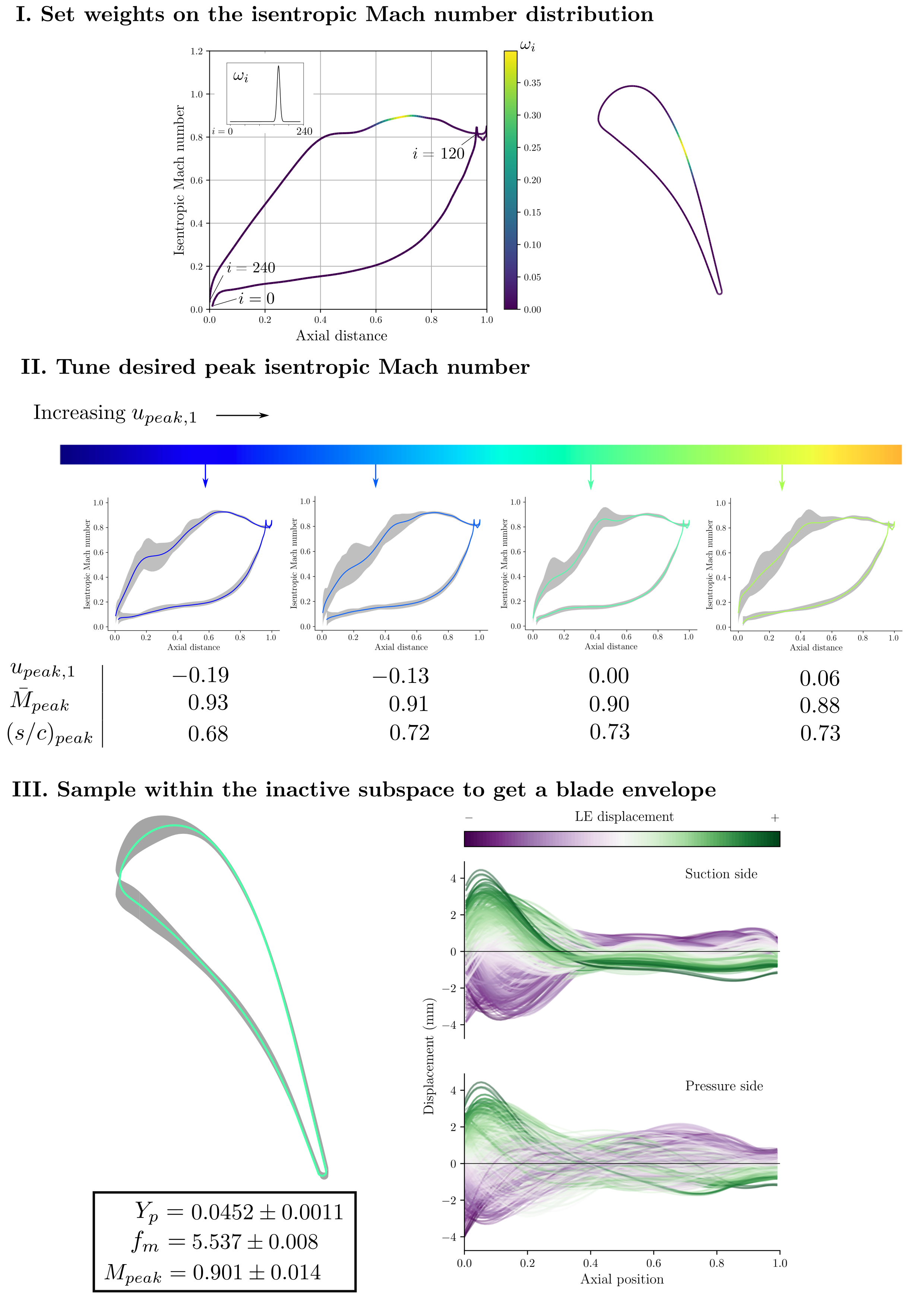}
\caption{Conditional tuning of the peak isentropic Mach number.}
\label{fig:inv_design_peak2}
\end{figure*}

Our goal is to extract a small number of active coordinates which capture the variation of the peak isentropic Mach number. Following the recipe in \Cref{sec:vecdr}, the gradient covariance matrix for the isentropic Mach number distribution $\mH_{peak}$ is required, where the weight matrix $\mR_{peak}$ is set such that diagonal elements corresponding to the peak are large.  This is illustrated in Step I in \cref{fig:inv_design_peak2}, where the color indicates the magnitude of the corresponding weights. To compute the necessary gradients with respect to each nodal isentropic Mach number, quadratic models are trained at each node using uniformly sampled designs from the full design space. When evaluated on independent validation data, the models on nodes with non-zero weights all achieved an $R^2$ value above 0.99. Note that we set a non-zero weight on multiple nodes in the vicinity of the peak. As we change the height of the peak, it is observed that the chord-wise location of the peak shifts slightly. This effect is accommodated by smoothing out the weights to a small region near the peak.
\begin{figure}
\centering
\includegraphics[width=0.6\linewidth]{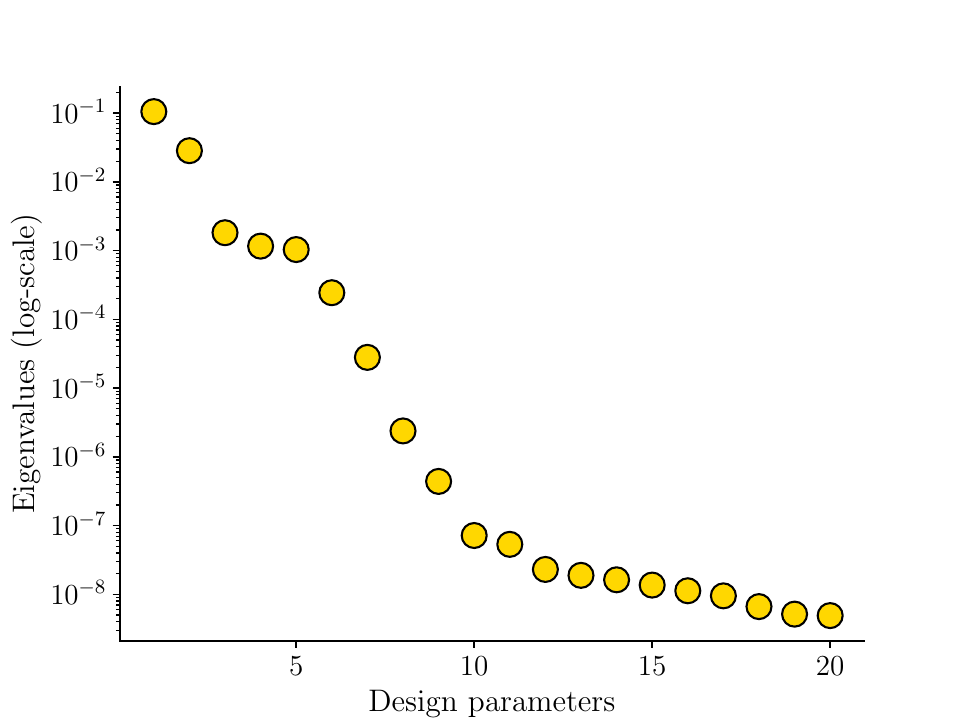}
\caption{Eigenvalues of $\mH_{peak}$, weighted for the peak isentropic Mach number.}
\label{fig:peak_eigs}
\end{figure}

Carrying out the eigendecomposition of $\mH_{peak}$,
\begin{equation}
\mH_{peak} = [\mW_{peak} \enskip \mV_{peak}] \begin{bmatrix} 
\mLambda_{peak,1} & \mathbf{0}\\
\mathbf{0} &\mLambda_{peak,2}
\end{bmatrix}
\begin{bmatrix}
\mW_{peak}^T\\
\mV_{peak}^T
\end{bmatrix}.
\end{equation}
The eigenvalues of $\mH_{peak}$ are shown in \cref{fig:peak_eigs}. The active coordinates can be defined using the first two eigenvectors for approximate control over the peak isentropic Mach number, namely,
\begin{align}
\begin{split}
\vu_{peak} &= \begin{bmatrix}
u_{peak,1} & u_{peak,2}
\end{bmatrix}^T \\
&=\mW_{peak}^T \vx.
\end{split}
\end{align}
This set of active coordinates can be combined with those corresponding to the loss coefficient and mass flow to form the following vector
\begin{equation}
\vu = \begin{bmatrix}
u_{loss}~ & u_{fm}~& u_{peak,1}~ & u_{peak,2}
\end{bmatrix}^T
\end{equation}
To control the peak isentropic Mach number while keeping the loss and mass flow constant, the procedure described in \Cref{sec:inverse_design_method} can be used. In this example, we only tune the first active coordinate $u_{peak,1}$ for illustrative purposes. The effect of changing this variable is shown in Step II in \cref{fig:inv_design_peak2}. Increasing $u_{peak,1}$ results in the flattening of the mean isentropic Mach number distribution. The location of the peak of the mean isentropic Mach number distribution moves gradually towards the trailing edge. If the coordinate is increased further, the mean peak shifts to the junction at the start of the plateau near mid-chord. The uncertainty bands around the distributions indicate the variation in the isentropic Mach number distribution for geometries with the same values of $\vu$. It should be noted that these uncertainty bands imply that some designs may have an earlier peak than predicted by the mean. Tighter control on the shape and location of the peak can be achieved by increasing the number of active coordinates, or using a more diffuse set of weights on the $\mR$ matrix. 

After setting appropriate values for the active coordinates, the hit-and-run sampling algorithm introduced in Part I can be used to sample designs with these coordinates. A blade envelope can be formed from these samples, following the same procedure as the previous examples. Step III in \cref{fig:inv_design_peak2} shows the blade envelope with $\vu = \mathbf{0}$. Compared with the previous blade envelopes, the tolerance on the profile from mid-chord to the trailing edge is reduced, especially on the suction side. Using the sample profiles, the tolerance covariance matrix and ensemble mean can be calculated as in Part I to arrive at an automatic decision criterion for the scrapping of manufactured blades, taking into account all three objectives.

\subsection{Setting the leading edge isentropic Mach number}
\begin{figure*}
\centering
\includegraphics[width=0.8\linewidth]{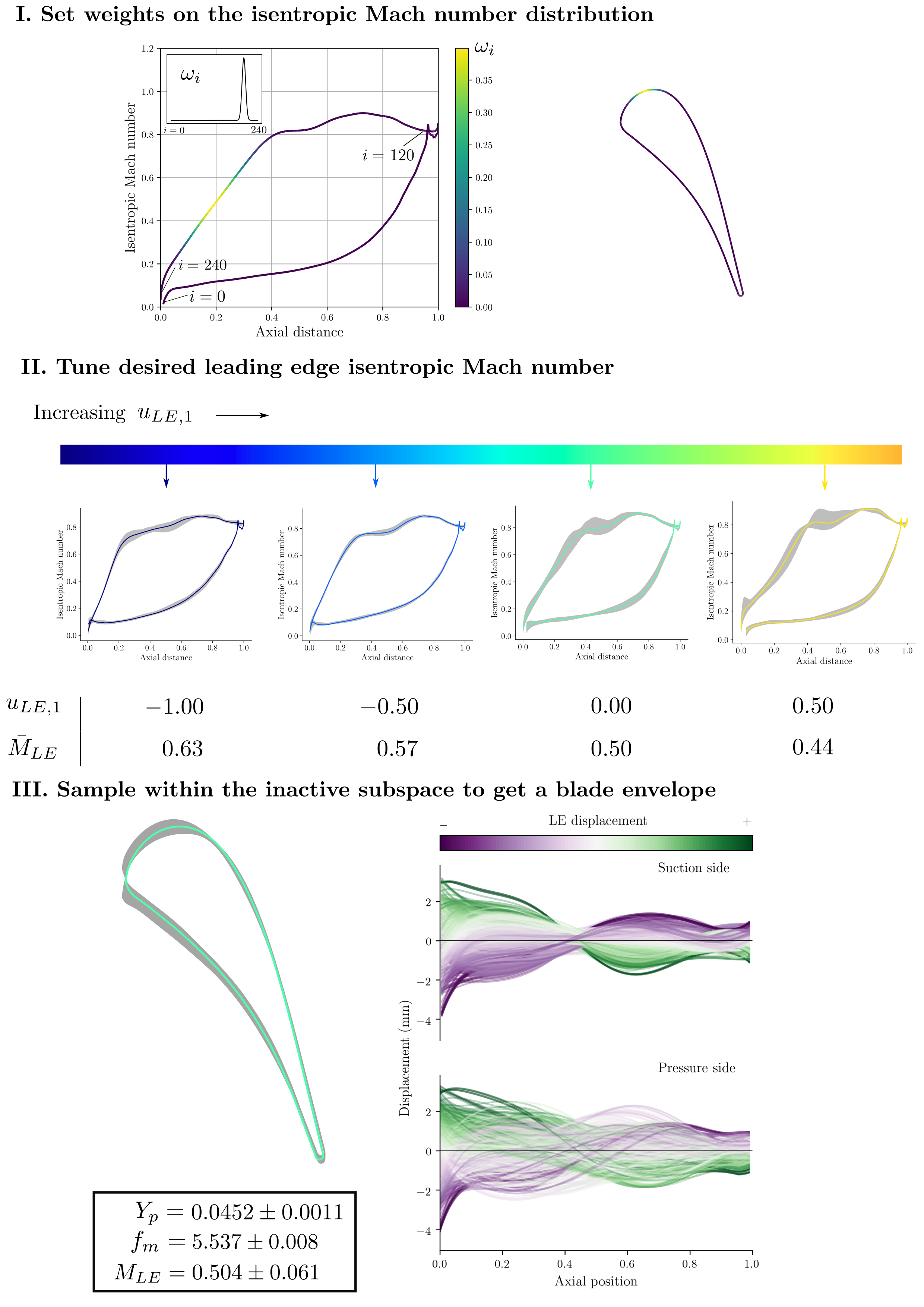}
\caption{Conditional tuning of the leading edge isentropic Mach number.}
\label{fig:inv_design_LE}
\end{figure*}

\begin{figure}
\centering
\includegraphics[width=0.6\linewidth]{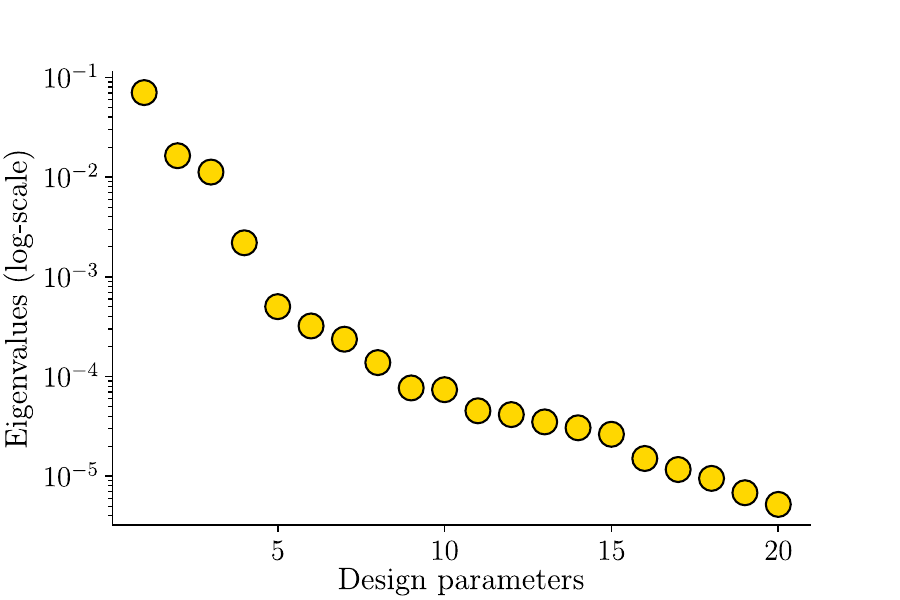}
\caption{Eigenvalues of $\mH_{LE}$, weighted for the leading edge isentropic Mach number.}
\label{fig:LE_eigs}
\end{figure}

The procedure can be repeated for a different setting of the weights on the isentropic Mach number distribution. Setting the weights as in Step I of \cref{fig:inv_design_LE} with a distribution centered around 20\% of the chord, the focus is now placed on the isentropic Mach number characteristics when the flow is accelerating before the peak on the suction side. Although this region of acceleration can span a significant portion of the chord including that near the leading edge, we call it the LE isentropic Mach number for brevity. Carrying out the eigendecomposition of the vector gradient covariance matrix $\mH_{LE}$,
\begin{equation}
\mH_{LE} = [\mW_{LE} \enskip \mV_{LE}] \begin{bmatrix} 
\mLambda_{LE,1} & \mathbf{0}\\
\mathbf{0} &\mLambda_{LE,2}
\end{bmatrix}
\begin{bmatrix}
\mW_{LE}^T\\
\mV_{LE}^T
\end{bmatrix},
\end{equation}
the spectrum shown in \cref{fig:LE_eigs} is obtained. Choosing the first four eigenvectors as the basis for the active subspace $\mW_{LE}$, the active coordinates are specified as
\begin{align}
\begin{split}
\vu_{LE} &= \begin{bmatrix}
u_{LE,1}~ & u_{LE,2}~ & u_{LE,3}~ & u_{LE,4}
\end{bmatrix}^T \\
&=\mW_{LE}^T \vx.
\end{split}
\end{align}
Concatenating this set of coordinates with the active coordinates of the loss and mass flow, we get
\begin{equation}
\vu = \begin{bmatrix}
u_{loss}~ & u_{fm}~& u_{LE,1}~ & u_{LE,2}~ & u_{LE,3}~ & u_{LE,4}
\end{bmatrix}^T.
\end{equation}
The effect of tuning the first active coordinate of the leading edge isentropic Mach number $u_{LE,1}$ is shown in Step II of \cref{fig:inv_design_LE}. As the coordinate is increased, the isentropic Mach number is reduced towards the leading edge, and the acceleration is milder and more distributed towards mid-chord. Step III shows the blade envelope obtained with samples setting $\vu = \mathbf{0}$. The geometric tolerance near the leading edge is tightened, especially on the suction side. 

\section{PRACTICAL UTILITY AND DISSEMINATION}
Thus far in this two-part paper, we have enumerated the core computational techniques required for generating a blade envelope under multiple constraints. We have articulated its utility in design (both forward and inverse) by identifying and exploiting the inactive subspace. Although all our efforts have focused on the LS89, the concept of a blade envelope extends well beyond any given airfoil, and indeed an isolated 2D cross-section. We envision blade envelopes forming a critical part of the design stage, facilitating discussions on trade-offs and tolerances. 

Beyond design, a key contribution of our work is a machine learning technique for deciding whether a manufactured blade's tolerance variations are permissible or not. This program can be easily incorporated within an existing coordinate measuring machine protocol and used in manufacturing assembly lines, with the aim of moving away from existing broad spectrum tolerances and opting for a more bespoke, design-centric one, which may offer opportunities for relaxing tolerances---thereby reducing manufacturing costs. Additionally, we see tremendous utility of the blade envelope in health monitoring; a topic we address in detail in a forthcoming paper. When inspecting blades that have gone through several hundred hours of operation, using either detailed GOM scans or borescope images, we can use the blade envelope to query whether a given blade needs to be refurbished or not.

The one high-level challenge that remains is how, in practice, the design, manufacturing and health monitoring teams cohesively converge and take advantage of a blade envelope, given that these teams are seldom collocated and may even be in different continents. To address this, we offer a simple plug-in from equadratures (used for doing the computational heavy lifting) \cite{seshadri2017effective} to a 3D modelling software package (we opt for Blender \cite{blender} given its open source nature). The latter can be easily adapted to industrial computer aided design (CAD) workflows.

\begin{figure}
\centering
\includegraphics[width=0.8\linewidth]{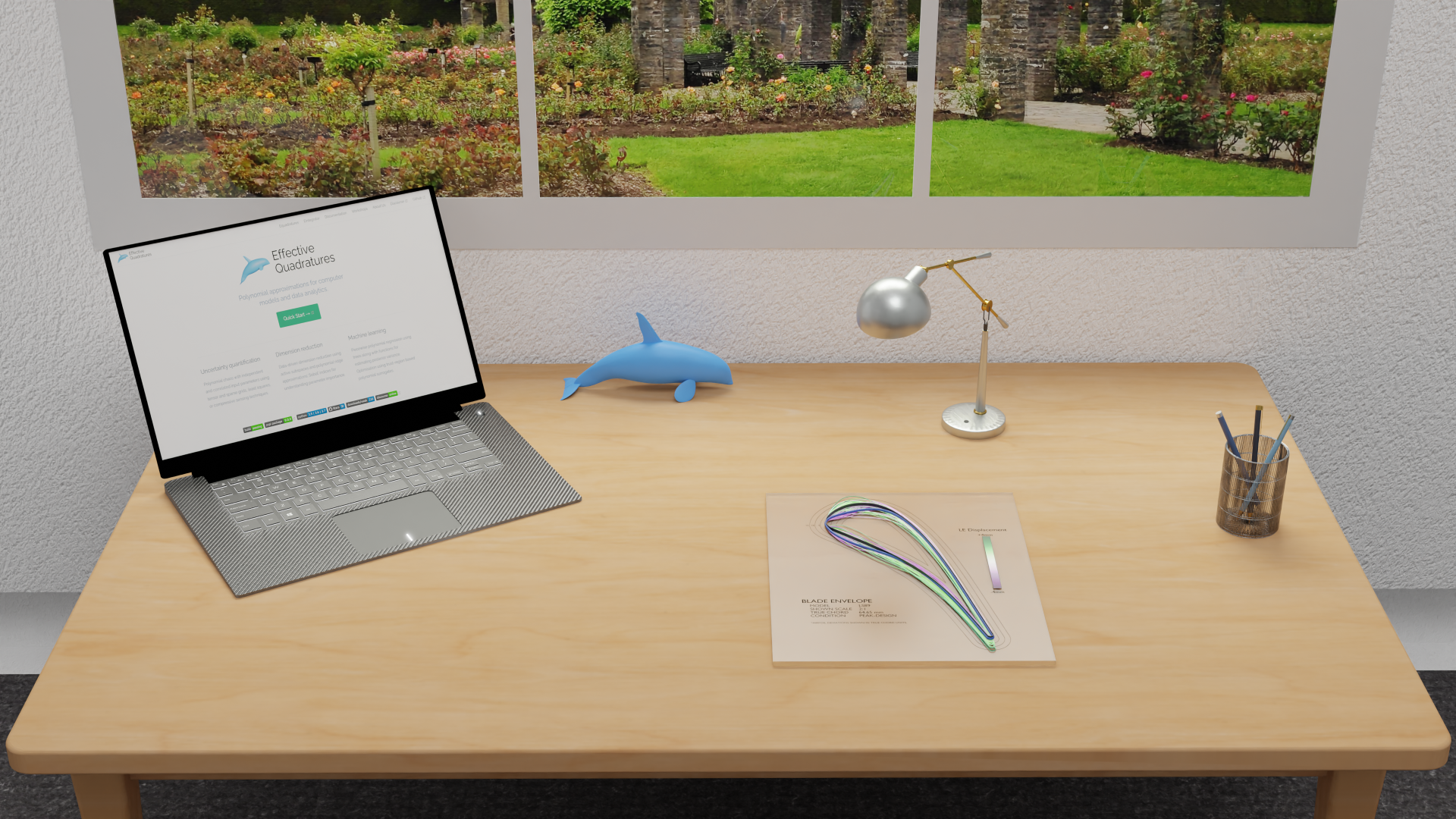}
\caption{The blade envelope from Figure~\ref{fig:cad_3} can be suitably scaled for 3D printing and placed on desktops.}
\label{fig:cad_4}
\end{figure}

\begin{figure*}
\centering
\includegraphics[width=0.8\linewidth]{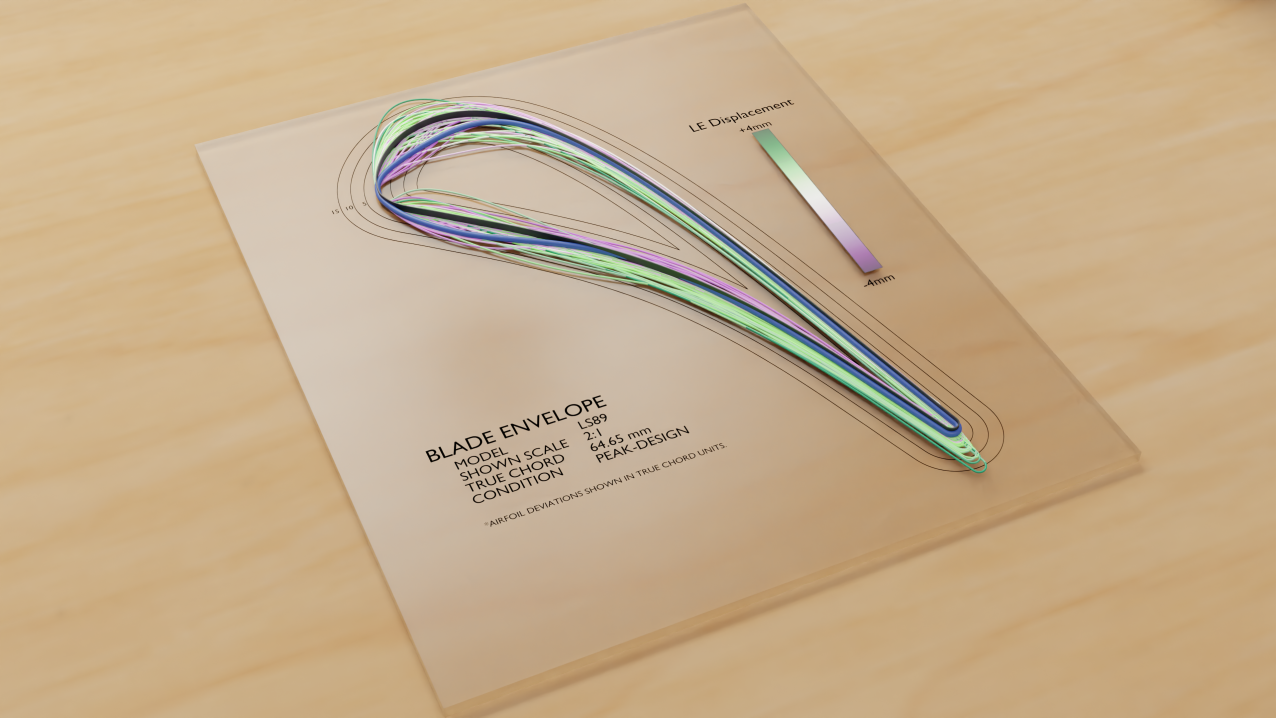}
\caption{An isometric view of the blade envelope in Figure~\ref{fig:cad_3}.}
\label{fig:cad_1}
\end{figure*}

\begin{figure*}
\centering
\includegraphics[width=0.8\linewidth]{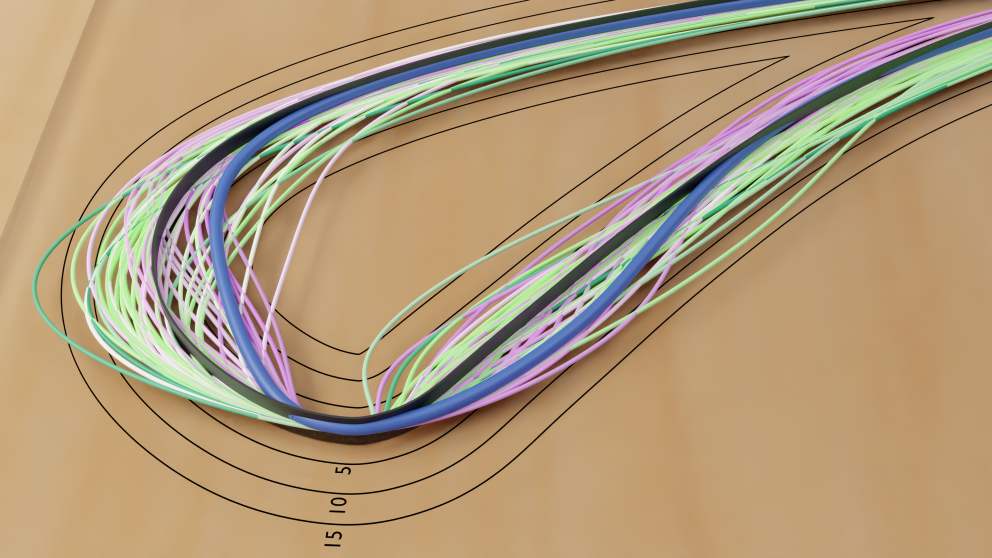}
\caption{Close-up of the leading edge of the blade envelope in Figure~\ref{fig:cad_3}. The demarcated iso-contour lines denote displacements at 5, 10 and 15 mm away from the nominal blade.}
\label{fig:cad_2}
\end{figure*}

We present a CAD representation of the controlled loss and mass flow rate blade envelope, shown in Figure~\ref{fig:cad_3}, in the Introduction. The purple and green airfoils correspond to profiles with negative and positive leading edge displacement respectively, just as plotted before, but now superimposed with the nominal blade (shown in black). Example invariant geometries can be highlighted for inspection, such as the one shown in blue here. Our objective in showing these images is to demonstrate how these envelopes, once frozen for the objectives desired, can be 3D printed---either at scale or as a scaled-up model---and can sit on the desktops of designers, manufacturing engineers, and engine maintenance inspectors (Figure~\ref{fig:cad_4}). With these 3D visualizations showing displacements to scale, we gain insights to numerous queries related to the manufacturing process, a few of which are captured below.
\begin{enumerate}
\item Guiding design trade-offs, such as the feasibility of increasing trailing edge thickness without loss of performance.
\item Informing engineers where to focus scanning efforts via the control zone. 
\item Comparing the tolerance implications across multiple nominal designs with different corresponding blade envelopes.
\end{enumerate}
These models can easily be transported, or indeed printed at different sites, and used to foster discussions on design and tolerance decisions. For completeness, close-up and isometric views of this blade envelope are shown in Figures~\ref{fig:cad_1} and \ref{fig:cad_2}. 

\section{CONCLUSIONS}
This paper extends the blade envelope framework to accommodate multiple objectives. These objectives can either be scalar-valued or a distribution defined on a surface. To take multiple objectives into consideration, an approach based on the intersection of inactive subspaces is described. For profile objectives, we propose a method based on vector-valued dimension reduction to seek an inactive subspace based on weights placed on different parts of the profile. Combining the inactive subspaces of the loss coefficient, mass flow function and the surface isentropic Mach number distribution, we demonstrated use of these methods to derive a blade envelope that delineates a boundary for designs that satisfy the constraints placed on all three objectives.

The collection of active coordinates corresponding to multiple objectives provide independent control over multiple objectives, by changing the active coordinates of one objective and keeping others constant. This is demonstrated through tuning the peak and leading edge isentropic Mach numbers within the inactive subspace of the loss and mass flow.

\section*{Acknowledgements}
The first author acknowledges financial support from the Cambridge Trust, Jesus College, Cambridge, and the Data-Centric Engineering programme of The Alan Turing Institute. The second author was funded through a Rolls-Royce research fellowship and the third author was funded through the Digital Twins in Aeronautics grant as part of the Strategic Priorities Fund EP/T001569/1.  The fifth author was supported by the Lloyd's Register
Foundation Programme on Data Centric Engineering and by The Alan Turing Institute under the
EPSRC grant EP/N510129/1. Thanks are due to  R\'{a}ul V\'{a}zquez for his support of this work. We also thank the anonymous reviewers for their comments, which helped improve the quality of the manuscript.

\bibliography{TURBO-20-1444}

\begin{thebibliography}{10}

\bibitem{wong_blade_nodate}
Wong, C.~Y., Seshadri, P., Scillitoe, A., Duncan, A., and Parks, G.~T.
\newblock ``Blade {Envelopes} {Part} {I}: {Concept} and {Methodology}''.
\newblock {\em Submitted to Journal of Turbomachinery}.

\bibitem{arts_aero-thermal_1992}
Arts, T., and Lambert~de Rouvroit, M., 1992.
\newblock ``Aero-{Thermal} {Performance} of a {Two}-{Dimensional} {Highly}
  {Loaded} {Transonic} {Turbine} {Nozzle} {Guide} {Vane}: {A} {Test} {Case} for
  {Inviscid} and {Viscous} {Flow} {Computations}''.
\newblock {\em Journal of Turbomachinery, {\bf 114}}(1), Jan., pp.~147--154.

\bibitem{torreguitart_optimization_2018}
Torreguitart, S., Verstraete, T., and Mueller, L., 2018.
\newblock ``Optimization of the {LS}89 {Axial} {Turbine} {Profile} {Using} a
  {CAD} and {Adjoint} {Based} {Approach}''.
\newblock {\em International Journal of Turbomachinery, Propulsion and Power,
  {\bf 3}}(3), Sept., p.~20.

\bibitem{montanelli_gradient_2015}
Montanelli, H., Montagnac, M., and Gallard, F., 2015.
\newblock ``Gradient {Span} {Analysis} {Method}: {Application} to the
  {Multipoint} {Aerodynamic} {Shape} {Optimization} of a {Turbine} {Cascade}''.
\newblock {\em Journal of Turbomachinery, {\bf 137}}(9), Sept., p.~091006.

\bibitem{seshadri_yuchi_parks_shahpar_2020}
Seshadri, P., Yuchi, S., Parks, G., and Shahpar, S., 2020.
\newblock ``Supporting multi-point fan design with dimension reduction''.
\newblock {\em The Aeronautical Journal, {\bf 124}}(1279), p.~1371–1398.

\bibitem{seshadri_turbomachinery_2017}
Seshadri, P., Shahpar, S., Constantine, P.~G., Parks, G.~T., and Adams, M.,
  2017.
\newblock ``Turbomachinery {Active} {Subspace} {Performance} {Maps}''.
\newblock {\em Journal of Turbomachinery, {\bf 140}}(4), Apr., p.~041003.

\bibitem{clark_step_2019}
Clark, C.~J., 2019.
\newblock ``A {Step} {Towards} an {Intelligent} {Aerodynamic} {Design}
  {Process}''.
\newblock In {ASME} {Turbo} {Expo} 2019: {Turbomachinery} {Technical}
  {Conference} and {Exposition}, American Society of Mechanical Engineers
  Digital Collection.

\bibitem{golub2013matrix}
Golub, G.~H., and van Loan, C.~F., 2013.
\newblock {\em Matrix {Computations}}, fourth~ed.
\newblock Johns Hopkins University Press, Baltimore, Apr.

\bibitem{WONG2020113383}
Wong, C.~Y., Seshadri, P., Parks, G.~T., and Girolami, M., 2020.
\newblock ``Embedded ridge approximations''.
\newblock {\em Computer Methods in Applied Mechanics and Engineering, {\bf
  372}}, p.~113383.

\bibitem{zahm_gradient-based_2018}
Zahm, O., Constantine, P.~G., Prieur, C., and Marzouk, Y.~M., 2020.
\newblock ``Gradient-{Based} {Dimension} {Reduction} of {Multivariate}
  {Vector}-{Valued} {Functions}''.
\newblock {\em SIAM Journal on Scientific Computing, {\bf 42}}(1), Jan.,
  pp.~A534--A558.
\newblock Publisher: Society for Industrial and Applied Mathematics.

\bibitem{seshadri2017effective}
Seshadri, P., and Parks, G., 2017.
\newblock ``Effective-quadratures ({EQ}): Polynomials for computational
  engineering studies''.
\newblock {\em The Journal of Open Source Software, {\bf 2}}, pp.~166--166.

\bibitem{blender}
{Blender Online Community}, 2018.
\newblock {\em Blender - a 3D modelling and rendering package}.
\newblock Blender Foundation, Stichting Blender Foundation, Amsterdam.

\end{thebibliography}
\bibliographystyle{asmems4}

\end{document}